\documentclass[twocolumn,superscriptaddress,showpacs,oneside,prl]{revtex4}

\begin{document}

\title{Spatial reflection and renormalization group flow of
quantum many-body systems with matrix product state representation}

\author{Li-Xiang Cen}
\affiliation{ Department of Physics, Sichuan University, Chengdu
610065, China}
\author{Z. D. Wang}
\affiliation{Department of Physics and Center of Theoretical and
Computational Physics, University of Hong Kong, Pokfulam Road, Hong
Kong, China}

\begin{abstract} The property of quantum many-body systems under
spatial reflection and the relevant physics of renormalization group
(RG) procedure are revealed. By virtue of the matrix product state
(MPS) representation, various attributes for translational invariant
systems associated with spatial reflection are manifested. We
demonstrate subsequently a conservation rule of the conjugative
relation for reflectional MPS pairs under RG transformations and
illustrate further the property of the fixed points of RG flows.
Finally, we show that a similar rule exists with respect to the
target states in the density matrix renormalization group algorithm.
\end{abstract}


\maketitle

Exploration of quantum many-body systems, particularly the
translational invariant systems defined on lattices, is one of the
most important topics in quantum physics and statistical physics.
Nevertheless, to our observation, the intrinsic attribute of the
system under spatial reflection and the relevant physics have less
been disclosed so far. In particular, a question whether or not the
species of quantum many-body states with matrix product construction
possess inherently a reflection symmetry \cite {ostlund} is yet to
be answered unambiguously. Here, we are motivated to reveal various
categories of quantum many-body systems under spatial reflection and
to explore the related property under the renormalization group (RG)
procedure.

The RG theory, including the seminal proposal of real-space
renormalization by Wilson \cite{wilson} and its renewed development
of the density matrix renormalization group (DMRG) method
\cite{white}, is one conceptual pillar of quantum many-body physics
and particularly constitutes a key theoretical element to quantum
critical phenomena. A theoretical picture of the standard DMRG
algorithm could be formulated in terms of variational optimization
within the representation of matrix product states (MPSs)
\cite{ostlund,umod}. In fact, the generality of this mathematical
representation for quantum many-body states, incorporating with the
fact that the ground state of most quantum systems could be well
approximated by a low-dimensional MPS, accounts unambiguously for
the origin of the power of the DMRG algorithm. Recently, it was
indicated that the MPS representation has a close connection with
the concept developed in the field of quantum information, leading
to significant progress, e.g., algorithms for periodic boundary
conditions \cite{verstra}, finite temperature \cite{temp}, and
simulating quantum systems of real time evolution \cite{time}.
Meanwhile, it was shown that a general RG procedure can be
established upon the quantum state itself via MPS representation
with properly defined coarse-graining transformations \cite{frank}.
With respect to the Wilsonian RG scheme on Hamiltonians, this
proposal suggests a specific rescaling approach to realize the scale
separation for quantum many-body states.

The main contribution of this paper are as follows. Firstly, by
invoking the spatial reflection transformation, we show that apart
from the symmetric states, the translational invariant MPSs could
have different attributes, that is, locally inequivalent to their
reflectional counterparts or differing from their reflectional
counterparts only by local unitary transformations. Subsequently, we
show that the conjugative relation of the reflectional MPS pair is
preserved along the recurrent coarse-graining transformations. Thus
a rule on conservation of reflective relation for RG flows is
indicated and the property of the leading fixed points is further
investigated. Finally, we demonstrate elaborately that a similar law
exists with respect to the target states in the numerical DMRG
procedure regarding its variational nature of performance.

Let us begin with the notation of the one-dimension translational
invariant MPS:
\begin{equation}
|\Psi \rangle =\frac 1{\sqrt{\mathcal{W}}}\sum_{s_1,\cdots
,s_N=1}^dTr(A^{s_1}\cdots A^{s_N})|s_1,\cdots ,s_N\rangle ,  \label{mps}
\end{equation}
where the set of $D\times D$ matrices $\{A^s,s=1,\cdots ,d\}$
parameterize the $N$-spin state with the dimension $D\leq
d^{N/2}$. The normalization factor is obtained as
$\mathcal{W}=TrE^N$, where $E=\sum_{s=1}^d\bar{A}^s\otimes A^s$ is
the so-called transfer matrix with bar denoting complex
conjugation. We now introduce a new state defined by a spatial
reflection on $|\Psi \rangle $, that is, $|\Psi _{rfl}\rangle
\equiv \mathcal{P}_N|\Psi \rangle $ where $\mathcal{P}_N$ is the
parity operator for the $N$-body system depicted by the action
$\mathcal{P}_N|s_1\cdots s_N\rangle =|s_N\cdots s_1\rangle $. In
fact, for the present situation with site-independent matrices
$\{A^s\}$, the reflectional counterpart state $|\Psi _{rfl}\rangle
$ is just an MPS represented by the set of matrices
$\{A_{rfl}^s\}$ where $A_{rfl}^s\equiv (A^s)^T$ denotes the matrix
transposition of $A^s$. This can be easily seen from the equation
\begin{eqnarray}
|\Psi _{rfl}\rangle &=&\frac
1{\sqrt{\mathcal{W}}}\sum_{\{s_i\}}Tr(A^{s_1}
\cdots A^{s_N})|s_N,\cdots ,s_1\rangle
\label{conju} \\
&=&\frac 1{\sqrt{\mathcal{W}}}\sum_{\{s_i\}}Tr[(A^{s_1})^T\cdots
(A^{s_N})^T]|s_1,\cdots ,s_N\rangle .  \nonumber
\end{eqnarray}
Note that the transfer matrix of $|\Psi _{rfl}\rangle $ is related to the
one of $|\Psi \rangle $ simply by
\begin{equation}
E_{rfl}=\sum_s\bar{A}_{rfl}^s\otimes A_{rfl}^s=E^T,  \label{adjtra}
\end{equation}
which indicates that $E$ and $E_{rfl}$ have exactly the same
spectrum. This leads clearly to the fact that any MPS has the same
correlation length \cite {qpt,ladder} with its reflectional
counterpart. Furthermore, the overlap of two reflective MPSs is
worked out to be
\begin{equation}
\eta \equiv \langle \Psi |\Psi _{rfl}\rangle =\frac{Tr(E^{T_2})^N}{TrE^N},
\label{overlap}
\end{equation}
where $E^{T_2}\equiv \sum_s\bar{A}^s\otimes (A^s)^T$. Clearly, Eq.
(\ref{overlap}) suggests a sufficient criterion for an MPS with
reflection symmetry, that is, the specified matrix $E^{T_2}$ should
have the same spectrum structure with that of the matrix $E$.

The well-known multipartite states in quantum information, typically
the Greenberger-Horne-Zeilinger state \cite{GHZ}, the cluster state
\cite {cluster} and the MPS of Affleck-Kennedy-Lieb-Tasaki model
\cite{AKLT}, are shown to be symmetric under spatial reflection.
Consider the cluster state as an example. By noting that the state
has an MPS representation \{$A^1$=(\negthinspace {\tiny $
\begin{array}{ll}
0 & 0 \\
1 & 1
\end{array}
\!\!$}), $A^2$=(\negthinspace {\tiny $
\begin{array}{ll}
1 & -1 \\
0 & 0
\end{array}
\!\!$})\}, one can work out that the matrix $E^{T_2}$ has the same
spectrum with that of $E$, hence $\eta =1$ according to Eq.
(\ref{overlap}). More specifically, it is verified that
$E^{T_2}=(I\otimes X)E(I\otimes X^{-1})$, where the matrix
$X$=(\negthinspace {\tiny $
\begin{array}{ll}
-1 & 1 \\
1 & 1
\end{array}
\!\!$}) is an invertible transformation connecting the matrix $A^s$ and its
transposition: $(A^s)^T=XA^sX^{-1}$.

The translational invariant MPS without parity symmetry, as will be
shown below, exists in general. In fact, it is of interest to
further distinguish two distinct categories for the translational
invariant states, namely, those locally inequivalent to their
reflectional counterparts and those differing from their
reflectional counterparts only by local unitary transformations.
Specifically, let us consider a translational invariant MPS $|\Psi
\rangle $ represented by
\begin{equation}
A^1=\left[
\begin{array}{ll}
1 & 0 \\
0 & -1
\end{array}
\right] ,A^2=\left[
\begin{array}{ll}
0 & 1 \\
0 & 0
\end{array}
\right] ,A^3=\left[
\begin{array}{ll}
0 & 0 \\
1 & g
\end{array}
\right] .  \label{example}
\end{equation}
It is shown that the corresponding transfer matrix $E$ has
eigenvalues ${-1,-1,(2+g^2\pm } \sqrt{4+g^4})/2$ that are distinctly
different from those of the matrix $E^{T_2}$ (except the case of
$g=0$). Therefore the specified reflectional MPS pair $|\Psi \rangle
$ and $|\Psi _{rfl}\rangle $ are different according to Eq.
(\ref{overlap}). Moreover, it can be shown that the two MPSs $|\Psi
\rangle $ and $|\Psi _{rfl}\rangle $ possess distinct correlation
features \cite{note0}, hence belong to different equivalence
classes, i.e., $|\Psi _{rfl}\rangle \neq U\otimes \cdots \otimes
U|\Psi \rangle $ where $U$ stands for local unitary transformations.

For the second example we consider an MPS $|\Psi \rangle $ with
\begin{equation}
A^1=\left[
\begin{array}{ll}
1 & 0 \\
0 & g
\end{array}
\right] ,A^2=\left[
\begin{array}{ll}
0 & 1 \\
0 & 0
\end{array}
\right] ,A^3=\left[
\begin{array}{ll}
0 & 0 \\
g & 0
\end{array}
\right] .  \label{example2}
\end{equation}
Explicitly, the corresponding transfer matrix $E$ has eigenvalues
${1+g^2,g,g,0}$ and the matrix $E^{T_2}$ has $2g,1,{g^2,0}$,
respectively. Therefore the MPS $|\Psi \rangle $ is different from
its reflectional counterpart $|\Psi _{rfl}\rangle $ in view that the
overlap between them is less than unity (apart from two exception
points of $g=\pm 1$). Interestingly, in this case the MPSs $|\Psi
\rangle $ and $|\Psi _{rfl}\rangle $ differ only by a local unitary
transformation, i.e., $|\Psi _{rfl}\rangle =U\otimes \cdots \otimes
U|\Psi \rangle $ where $U=|1\rangle \langle 1|+|2\rangle \langle
3|+|3\rangle \langle 2|$. Definitely, the present example suggests a
special category of translational invariant states that relate to
their reflectional counterparts by non-trivial local unitary
transformations. In general, the representative matrices of this
special sort of MPSs satisfy $(A^i)^T=\sum_jU_j^i(XA^jX^{-1})$,
where $U_j^i$ is the representative matrix accounting for the local
unitary transformation and $X$ is an invertible matrix, say, it is
obtained as $X$=(\negthinspace {\tiny $
\begin{array}{ll}
g & 0 \\
0 & 1
\end{array}
\!\!$}) for the present case of Eq. (\ref{example2}).

Now let us consider the relation of the reflectional MPS pair under
the RG transformation. Following Ref. \cite{frank}, to perform the
coarse-graining procedure for the state $|\Psi \rangle $ in Eq.
(\ref{mps}), one needs firstly merge the representative matrices for
neighboring sites $\tilde{A}^{(pq)}=A^pA^q$. Then an appropriate
representative for the equivalence class can be selected out via the
singular value decomposition
\begin{equation}
\tilde{A}_{(\alpha \beta )}^{(pq)}=\sum_{l=0}^{d^{\prime }}U_l^{(pq)}
\lambda _lV_{(\alpha \beta )}^l,  \label{RG}
\end{equation}
where $(pq)$ and $(\alpha \beta )$ are understood as combined
indices, and $d^{\prime }\leq \min \{D^2,d^2\}$ denotes the number
of non-zero singular values of the matrix $\tilde{A}_{(\alpha \beta
)}^{(pq)}$. The state after one-step RG transformation can therefore
be characterized by the new representative matrices
\begin{equation}
A^p\rightarrow A^{^{\prime }l}=\lambda _lV^l.  \label{RGres}
\end{equation}

Consider now the specified RG performance on the reflectional
counterpart state $|\Psi _{rfl}\rangle $ represented by
$\{A_{rfl}^p\}$. In view of the relation of the coarse-grained
matrices $\tilde{A}_{rfl}^{(pq)}\equiv
A_{rfl}^pA_{rfl}^q=(\tilde{A}^{(qp)})^T$, one has
\begin{equation}
(\tilde{A}_{rfl})_{(\alpha \beta )}^{(pq)}=\tilde{A}_{(\beta \alpha
)}^{(qp)}=\sum_{l=1}^{d^{\prime }}U_l^{(qp)} \lambda _lV_{(\beta \alpha )}^l.
\label{RGad}
\end{equation}
Hence the RG transformation on $|\Psi _{rfl}\rangle $ gives rise to
\begin{equation}
A_{rfl}^p\rightarrow A_{rfl}^{\prime l}=\lambda _l(V^l)^T=(A^{\prime l})^T.
\label{RGadjre}
\end{equation}
Clearly, Eqs. (\ref{RGres}) and (\ref{RGadjre}) show that the
relation of spatial reflection is preserved for the reflectional MPS
pair under the RG transformation. In fact, the recurrent RG
performance indicated by Eqs. (\ref{RG})-(\ref{RGadjre}) suggests an
intriguing conjugative structure of RG flow for the translational
invariant states. This special flow configuration will continue
along the RG procedure until the states reach their fixed points.
Furthermore, since the corresponding transfer matrices after
one-step RG transformation are given by $E^{\prime }=E^2$ and
$E_{rfl}^{\prime }=(E^T)^2=(E^{\prime })^T$, the overlap of the
reflectional MPS pair under the recurrent RG performance is obtained
explicitly as
\begin{equation}
\eta \rightarrow \eta ^{\prime }=\frac{Tr[(E^2)^{T_2}]^{N/2}}{TrE^N}
\rightarrow \cdots \rightarrow \frac{Tr[(E^\infty )^{T_2}]^n}{TrE^N },
\label{overlap1}
\end{equation}
where $n=N/\chi $ and we have denoted by $E^\infty $ and
$E_{rfl}^\infty =(E^\infty )^T$ the transfer matrices of the two
reflectional fixed points, i.e., $E^\infty \equiv
\lim_{\chi\rightarrow\infty }E^\chi $.

The above depicted conservation law of reflective relation for RG
flows is applicable for both the two category MPSs: those $|\Psi
\rangle $ differing from $|\Psi _{rfl}\rangle $ by local unitary
transformations and those $|\Psi \rangle $ locally inequivalent to
$|\Psi _{rfl}\rangle $. For the former case, although $|\Psi \rangle
$ and $|\Psi _{rfl}\rangle $ are viewed to be equivalent under the
coarse-graining transformation, the attribute of spatial reflection
is retained along the RG procedure even at the fixed point. In
detail, let us examine the MPS of Eq. (\ref{example2}). It is direct
to calculate that the two reflectional fixed points are
characterized by $E^\infty =|\Phi _R\rangle \langle \Phi _L|$ and
$E_{rfl}^\infty =|\Phi _L\rangle \langle \Phi _R|$, where $|\Phi
_R\rangle =(|00\rangle +g^2|11\rangle )/(1+g^2)$ and $|\Phi
_L\rangle =|00\rangle +|11\rangle $. The corresponding
representative matrices of fixed point MPSs $|\Psi ^\infty \rangle $
and $|\Psi _{rfl}^\infty \rangle $ are obtained respectively as
\begin{equation}
\{A_\infty^s\}=\left\{ \left[
\begin{array}{ll}
1 & 0 \\
0 & 0
\end{array}
\right] ,\left[
\begin{array}{ll}
0 & g \\
0 & 0
\end{array}
\right] ,\left[
\begin{array}{ll}
0 & 0 \\
1 & 0
\end{array}
\right] ,\left[
\begin{array}{ll}
0 & 0 \\
0 & g
\end{array}
\right] \right\}   \label{fixed2}
\end{equation}
and $\{(A_\infty^{rfl})^s=(A_\infty^s)^T,s=1,\cdots ,4\}$. It is
readily verified that $|\Psi _{rfl}^\infty \rangle =U\otimes \cdots
\otimes U|\Psi ^\infty\rangle $ where the local unitary
transformation $U=|1\rangle \langle 1|+|4\rangle \langle
4|+|2\rangle \langle 3|+|3\rangle \langle 2|$.

For the situation specified by Eq. (\ref {example}) in which $|\Psi
\rangle $ and $|\Psi _{rfl}\rangle $ possess different correlation
feature, the corresponding fixed points could be obtained similarly.
In detail, since there is no degeneracy in the largest eigenvalue of
the transfer matrix $E$, the fixed points are characterized, up to
an irrelevant normalization factor, by $E^\infty =|\Phi _R\rangle
\langle \Phi _L|$ and $E_{rfl}^\infty =|\Phi _L\rangle \langle \Phi
_R|$, where
\begin{eqnarray}
|\Phi _R\rangle  &=&|00\rangle +\frac{\kappa +g^2}2|11\rangle
~~(\kappa
\equiv \sqrt{4+g^4}),  \label{eigenv} \\
|\Phi _L\rangle  &=&|00\rangle +\frac g{\kappa +1-g^2}(|01\rangle
+|10\rangle )+\frac{\kappa +g^2}2|11\rangle . \nonumber
\end{eqnarray}
One can verify from Eq. (\ref{overlap}) that for the corresponding
fixed point states $|\langle \Psi ^\infty |\Psi _{rfl}^\infty
\rangle |<1$. Notably, it turns out that the state $|\Psi ^\infty
\rangle $ differs from $|\Psi _{rfl}^\infty \rangle $ only by a
local unitary transformation \cite{note1}. Physically, it is
understood that all correlation functions decay exponentially along
the RG flow and become zero at the fixed point. Therefore the
attribute of the fixed point described above is a general feature
one exactly expects.

So far, we have revealed various attributes and the relevant physics
of RG flows for translational invariant MPSs under spatial
reflection. Now let us consider the DMRG scheme on the specified
lattice system. Note that for a system without parity symmetry,
i.e., $H_{rfl}\equiv\mathcal{P}(H)\neq H$, there is no reflection
relation between the system block and the environment block in the
DMRG algorithm any more. On the other hand, it is obvious that the
systems $H$ and $H_{rfl}$ have corresponding exact ground states
related by the reflection transformation. Hence it is interesting to
explore whether the performance of DMRG algorithm could warrant the
reflection relation between target states of reflective systems.

In detail, let us look into the DMRG procedure with $\mathbf{B\bullet
\bullet B}$ configuration for one-dimensional spin chains. The standard DMRG
iterative performance could be described as follows. Suppose that the
superblock comprises two blocks and two spins in between at a certain step.
The system block $\mathbf{B}_L$ contains spins $1,\cdots M-1$, and the
environment block $\mathbf{B}_R$ contains $M+1,\cdots ,2M-1$. The states of
two spins in between are denoted as $|s_M\rangle $ and $|s_M^{\prime
}\rangle $, respectively. The target state, i.e., the ground state of the
superblock has the following form
\begin{equation}
|\Psi \rangle =\sum_{s_M,s_M^{\prime }=1}^d\sum_{\alpha ,\beta
=1}^DA_{\alpha ,\beta }^{s_M,s_M^{\prime }}|\alpha \rangle
_{M-1}^L|s_M\rangle |s_M^{\prime }\rangle |\beta \rangle _{M-1}^R,
\label{superblock}
\end{equation}
where the orthonormal bases $|\alpha \rangle _{M-1}^{L,R}$ are
obtained from previous steps and the tensor $A_{\alpha ,\beta
}^{s_M,s_M^{\prime }}$ is determined such that the target state
minimizes the energy. From Eq. (\ref {superblock}), the reduced
density matrices of the left and right half superblock,
$\mathbf{B}_L\bullet $ and $\bullet \mathbf{B}_R$, are derived
directly by virtue of the following singular value decomposition
\begin{eqnarray}
A_{\alpha ,\beta }^{s_M,s_M^{\prime }} &=&(U\Sigma V)_{(s_M,\alpha
),(s_M^{\prime },\beta )}  \nonumber \\
&=&\sum_{\alpha ^{\prime }=1}^{d\times D}U_{(s_M,\alpha ),\alpha ^{\prime
}}\Sigma _{\alpha ^{\prime }}V_{\alpha ^{\prime },(s_M^{\prime },\beta )},
\label{SVD}
\end{eqnarray}
where $(s_M,\alpha )$ and $(s_M^{\prime },\beta )$ are understood as
combined indices and $\Sigma $ is a diagonal matrix with elements
(singular values) $\Sigma _{\alpha ^{\prime }}$, sorted in
decreasing order, accounting for square roots of eigenvalues of the
reduced density matrices. Then a truncation algorithm to achieve new
system and environment blocks for the next step iteration,
$\mathbf{B}_L\bullet \rightarrow \mathbf{B}_L^{\prime }$ and
$\bullet \mathbf{B}_R\rightarrow \mathbf{B}_R^{\prime }$, is
performed by retaining only following $D$ eigenvectors with the
largest eigenvalues
\begin{eqnarray}
|\alpha ^{\prime }\rangle _M^L &=&\sum_{s_M=1}^d\sum_{\alpha =1}^DU_{\alpha
,\alpha ^{\prime }}^{[M],s_M}|\alpha \rangle _{M-1}^L\otimes |s_M\rangle ,
\nonumber \\
|\beta ^{\prime }\rangle _M^R &=&\sum_{s_M=1}^d\sum_{\beta =1}^DV_{\beta
^{\prime },\beta }^{[M],s_M^{\prime }}|\beta \rangle _{M-1}^R\otimes
|s_M^{\prime }\rangle .  \label{recu}
\end{eqnarray}
Here, $U_{\alpha ,\alpha ^{\prime }}^{[M],s_M}$ and $V_{\beta
^{\prime },\beta }^{[M],s_M^{\prime }}$ ($\alpha ^{\prime },\beta
^{\prime }=1,\cdots ,D$) are just sub-unitary matrices truncated
from $U_{(s_M,\alpha ),\alpha ^{\prime }}$ and $V_{\beta ^{\prime
},(s_M^{\prime },\beta )}$, respectively, and they fulfill the
relation
\begin{eqnarray}
I &=&\sum_{s_M}(U^{[M],s_M})^{\dagger }U^{[M],s_M}  \nonumber \\
&=&\sum_{s_M^{\prime }}V^{[M],s_M^{\prime }}(V^{[M],s_M^{\prime
}})^{\dagger }.  \label{relation}
\end{eqnarray}
In terms of the MPS representation, the target state in the above DMRG
iteration procedure can be depicted distinctly as
\begin{eqnarray}
|\Psi \rangle &=&\sum_{\{s_i\}}Tr(U^{[1],s_1}\cdots
U^{[M-1],s_{M-1}}A^{s_M,s_M^{\prime }}  \label{MPSPBC} \\
&&\times V^{[M-1],s_{M+1}}\cdots V^{[1],s_{2M-1}})|s_1\cdots s_{2M-1}\rangle
,  \nonumber
\end{eqnarray}
where we have used the notation $|s_1\cdots s_{2M-1}\rangle
=|s_1\cdots s_M,s_M^{\prime }\cdots s_{2M-1}\rangle $ and the
summation indices $\{s_i\}$ run over all the $2M$ spins. Note that
the MPS here is site dependent, i.e., no longer translational
invariant, and we have adopted periodic boundary conditions in Eq.
(\ref{MPSPBC}). The DMRG procedure is now clearly phrased as that
once the transformation $A^{s_M,s_M^{\prime }}\rightarrow
(U^{[M],s_M},V^{[M],s_M^{\prime }})$ is derived, then both system
and environment blocks increase in length by one site and the
algorithm is iterated until some desired final length is reached.

 The promising conjugated DMRG flow for the reflected
system $H_{rfl}$ is outlined below. It turns out that the target
state of the superblock for the system $H_{rfl}$ relates to the
original one (\ref{MPSPBC}) merely by an action of the parity
operator $|\Psi _{rfl}\rangle =\mathcal{P}|\Psi \rangle $. Namely,
one has
\begin{eqnarray}
|\Psi _{rfl}\rangle &=&\sum_{\{s_i\}}Tr(\tilde{V}^{[1],s_1}\cdots
\tilde{V}^{[M-1],s_{M-1}} \tilde{A}^{s_M,s_M^{\prime }}  \label{dmrgcon} \\
&&\times \tilde{U}^{[M-1],s_{M+1}}\cdots \tilde{U}^{[1],s_{2M-1}})|s_1\cdots
s_{2M-1}\rangle ,  \nonumber
\end{eqnarray}
where the tensors in the last expression are defined by
\begin{eqnarray}
\tilde{V}^{[k],s_k} &=&(V^{[k],s_k})^T,\tilde{U}
^{[k],s_{2M-k}}=(U^{[k],s_{2M-k}})^T,  \nonumber \\
\tilde{A}^{s_M,s_M^{\prime }} &=&(A^{s_M^{\prime },s_M})^T.  \label{transp}
\end{eqnarray}
To demonstrate that the reflecting forms (\ref{MPSPBC}) and
(\ref{dmrgcon}) of target states are preserved along the DMRG
iteration for systems $H$ and $H_{rfl}$, we need to prove (i) the
formulated states (\ref{MPSPBC}) and (\ref {dmrgcon}) minimize the
energy of the two reflected systems simultaneously; (ii) the
truncation algorithm of DMRG warrants that the resulted new
representative matrices and target states satisfy repetitiously
the indicated reflective relation.

Point (i) is readily verified since the expected values of Hamiltonians $H$
and $H_{rfl}$ over the states (\ref{MPSPBC}) and (\ref{dmrgcon}) satisfy
faithfully
\begin{equation}
E=\langle \Psi |H|\Psi \rangle /\mathcal{W} =\langle \Psi
_{rfl}|H_{rfl}|\Psi _{rfl}\rangle /\mathcal{W}   \label{varia}
\end{equation}
with the normalization factor $\mathcal{W}=\langle \Psi |\Psi
\rangle =\langle \Psi _{rfl}|\Psi _{rfl}\rangle $. To demonstrate
the point (ii), we note the following relation
\begin{eqnarray}
\tilde{A}_{\alpha ,\beta }^{s_M,s_M^{\prime }} &=&(V^T\Sigma
U^T)_{(s_M,\alpha ),(s_M^{\prime },\beta )}  \nonumber \\
&=&\sum_{\beta ^{\prime }=1}^{d\times D}V_{\beta ^{\prime },(s_M,\alpha
)}\Sigma _{\beta ^{\prime }}U_{(s_M^{\prime },\beta ),\beta ^{\prime }}.
\label{SVD2}
\end{eqnarray}
Consequently by virtue of the specified DMRG truncation
prescription one obtains the new representative matrices $V_{\beta
^{\prime },(s_M,\alpha )}\rightarrow V_{\beta ^{\prime },\alpha
}^{[M],s_M}=\tilde{V}_{\alpha ,\beta ^{\prime }}^{[M],s_M}$ and
$U_{(s_M^{\prime },\beta ),\beta ^{\prime }}\rightarrow U_{\beta
,\beta ^{\prime }}^{[M],s_M^{\prime }}=\tilde{U}_{\beta ^{\prime
},\beta }^{[M],s_M^{\prime }}$, and the corresponding recursive
relations [cf. Eq. (\ref{recu})]. This completes our proof that
the presented forms (\ref{MPSPBC}) and (\ref{dmrgcon}) of target
states are preserved along the DMRG iterative procedure for the
two reflective systems.

In conclusion, we have disclosed the property of quantum many-body
systems under spatial reflection and revealed a universal
conjugative flow structure for both the RG scheme on translational
invariant MPSs and the DMRG algorithm. An intriguing extension to
high spatial dimensions via projected entangled pair states
\cite{peps} is awaited for us to explore further.

This work was supported by the NSFC grants (10604043 and 10429401),
the RGC grants of Hong Kong (HKU7045/05P and HKU7051/06P).


\begin{references}
\bibitem{note0} The difference of the correlation feature between
$|\Psi \rangle $ and $|\Psi _{rfl}\rangle $ can be explicitly shown,
say, in view that the correlated density matrices $c\equiv
\rho_{123}-\rho_{12}\otimes\rho_3$ and $c^{rfl}\equiv
\rho^{rfl}_{123}-\rho^{rfl}_{12}\otimes\rho^{rfl}_3$ have different
spectrum.
\bibitem{note1} Equivalence between $|\Psi ^\infty \rangle $
and $|\Psi _{rfl}^\infty \rangle $ under local transformations could
be shown by the fact that there exists an invertible $X$ such that
$E_{rfl}^\infty =X\otimes \bar {X}E^\infty (X\otimes \bar
{X})^{-1}$.
\end{references}
\end{document}